# Evaluating the predicted eruption times of geysers in Yellowstone National Park

**Daniel J. Rhee[1], Ka Yee Yeung[2]**

[1]Interlake High School, Bellevue, Washington
[2]University of Washington Tacoma, Tacoma, Washington

**SUMMARY**

This study aims to evaluate the accuracy of predicted eruption times of popular geysers in the Yellowstone National Park.  The Yellowstone National Park was the first national park in the United States and is known for its geothermal features consisting of many highly popular geysers such as the Old Faithful. Geysers are fascinating to national park visitors because their eruptions could range from small bubbles to jets of water that are hundreds of meters high, and their eruptions could last from seconds to hours. To help tourists plan their visits, the US National Park Service and other independent groups publish predicted eruption times of popular geysers. We hypothesized that the models developed by the US National Park Service are very accurate with little discrepancy from independent analysis, as park rangers monitor the geysers constantly and likely adjust their models over time according to changing conditions underground, and patterns observed. In addition, since researchers in the park likely rely on these predictions, the models would need to be fine-tuned to ensure that no unnecessary effort or resources are wasted in probing the geysers for variables such as temperature and acidity. In this study, we focused on the Old Faithful and Beehive Geyser by downloading actual eruption times, conducting statistical regression analyses, studying the patterns of eruption times, and evaluating the accuracy of different statistical models.

**INTRODUCTION**

The Yellowstone National Park, established in 1872, was the first national park in the United States. One of the main attractions of Yellowstone is the hydrothermal vents such as geysers. There are more than 10,000 hydrothermal features at Yellowstone, with over 500 geysers making up around half of the world's geysers (1). Most of the geysers within Yellowstone reside in the Upper Geyser Basin (2). The Upper Geyser Basin is perhaps the most famous attraction of Yellowstone, including the famous Old Faithful, Grand and Castle geysers, with over 150 hydrothermal features in 1 square mile (1).

Geysers are pools of boiling hot water, around 93C at the surface, which are constricted near the top (3). The water that geysers expel are heated from Yellowstone's magma, as Yellowstone is an old volcano (3). Water is prevented from freely flowing and heat builds up below the surface, while the sheer weight of the rocks prevents the water from boiling (3). Eventually, the water reaches a point where the underground water forces the surface water out and water is expelled through the top, thus rapidly emptying the pool of water below ground. Eruptions may last from seconds to hours, with geysers such as the Steamboat Geyser reaching over 300 feet high (4).

Geothermal features such as the geysers are formed through a complex series of earthquakes that create deep underground piping which feeds water into the pools, before being heated up and expelled out of geysers, or boiled off in smaller bubblers. Since much of the water inflow is connected, it is hard to know when a geyser may have enough water to erupt, and the constantly evolving underground structures also change how fast the heating process is (5). Because of the extreme underground heat, it is near impossible to simply send autonomous mapping robots to determine the pathing of the pipes, which is why a vast majority of the Yellowstone underground piping remains unknown (6). Due to the complex and random nature of naturally formed piping, it is extremely difficult to predict the eruption time of geysers. The dynamics of geyser eruptions at Yellowstone have fascinated scientists for more than two centuries (7).

In this study, we focused on the Old Faithful and Beehive Geyser in the popular Upper Geyser Basin. The Old Faithful is one of the most famous geysers in the world, as well as one of the most visited. Old Faithful is so popular that the park runs off the "Old Faithful clock" where activities and programs are planned around the eruption of the geyser, due to its predictable nature and grand spectacle. This makes Old Faithful a very important geyser to generate precise predictions of eruption time. Beehive Geyser is not the largest or most grand geyser in the park; however, it is situated on the same boardwalk as Old Faithful, and can erupt more than once a day, making it a popular location for tourists (8). In addition, the Beehive Indicator erupts before the Beehive Geyser to indicate an eruption, which results in a different method of prediction to further diversify the methods of prediction. In addition, Beehive Geyser is part of a larger underground network of piping, unlike Old Faithful which is independently piped (9), adding another layer of complexity.

Very rarely are there geysers such as Old Faithful that erupt in highly predictable intervals, hence the name being "Old Faithful" (10). However, many geysers are predictable primarily due to a form of an indicator. Specifically, geysers such as the Beehive Geyser are predictable due to a second smaller geyser that spouts water prior to an eruption, and the Grand Geyser is predictable due to a boiling pool that bubbles prior to eruptions.

To help tourists plan their visits, the US National Park Service and other independent groups, such as the GeyserTimes app (11), that publish predicted eruption times of popular geysers. The goal of this study is to evaluate the accuracy of predicted eruption times of geysers at Yellowstone. We hypothesized that the models developed by the US National Park Service are very accurate with little discrepancy from independent analysis, as park rangers monitor the geysers constantly and likely adjust their models over time according to changing conditions underground, and patterns observed. In addition, since researchers in the park likely rely on these predictions, the models would need to be fine-tuned to ensure that no unnecessary effort or resources are wasted in probing the geysers for variables such as temperature and acidity.

**RESULTS**

Data consisting of eruption start time, duration of eruption, height of eruption, water temperature and pH were downloaded from GeyserTimes (11). Missing and incomplete data were included in this downloaded data. Therefore, complete datapoints for Old Faithful and Beehive Geyser were filtered using a Python script.

Old Faithful has been noted to exhibit correlation between the duration of an eruption and the time between eruptions (9). The relationship between the time between eruptions (y-axis) and the duration of the previous eruption (x-axis) is shown in **Figure 1**. Three types of regression methods, namely, linear, exponential, and logistic, were used to capture this observed relationship. Specifically, time of the next eruption is predicted using the duration of the previous eruption as the independent variable. To evaluate the accuracy of these models, the percentage of the time a prediction was within a certain range of the actual eruption time was calculated. **Table 1** shows the percentage of time each model accurately predicted the geyser within a certain time range, starting at 2 minutes (+/- 1 minute), and going up to a window of 30 minutes (+/- 15 minutes). Since a small time range implies high accuracy, the percentage of time each model accurately predicted the eruption time increases as the time range increases.

The National Park Service attempts to have 90% of the Old Faithful predictions fall within a 20-minute range of the predicted time as they have determined this is the correct balance of reliability and range (12). This 20-minute range is shown as (+/- 10 minutes) and in **bold** in **Table 1**. However, none of the regression methods nor the official predictions manage to have 90% accuracy within a 20-minute window. While the National Park Service's model does provide the most accurate predictions for windows up to 16-minutes (+/- 8 minutes), the sigmoidal model creates more accurate predictions beyond this time range. While the sigmoidal model is comparable to the National Park Service's model within the 20-minute window, it is significantly more accurate than the official model when the range increases. Therefore, our results confirmed our initial hypothesis that the models developed by the US National Park Service are very accurate is correct for windows up to 16-minutes (+/- 8 minutes).

Beehive Geyser was chosen as the next geyser for analysis, as it is situated near Old Faithful and is also known to be predictable (13). Beehive Geyser has an indicator known as Beehive's Indicator that creates a smaller eruption prior to the main eruption. This is because the indicator is used to relieve pressure preventing the main geyser eruption from erupting until the channel to the indicator is blocked. Different statistical methods were used to predict the eruption time of Beehive Geyser due to the existence of the indicator.

To measure if the time between the indicator and main eruption had changed over time, the difference between the eruption time of Beehive Geyser and the eruption time of Beehive Indicator was calculated for each eruption, and this difference was plotted against the order of eruption, as shown in **Figure 2**. Adding a trendline onto the graph revealed a gradual decrease of this time difference with less than a minute over 50 years, thus showing little change in this difference over time. Therefore, we assume a constant time interval between the eruption of the Beehive Indicator and the Beehive Geyser. This constant was estimated by computing the mean, median and mode of the differences between Beehive eruption times and indicator times across all eruptions to determine which method would be the best predictor of this constant. The mean was 13.3 minutes, the median was 13 minutes, and the mode was 14 minutes. These are in contrast to the 17 minutes adopted by the National Park Service (13). These estimated constants using the mean, median and mode were added to the Beehive Indicator eruption time to predict the Beehive Geyser eruption time.

**Figure 3** shows the percentage time that the actual Beehive eruption occurred within 10 minutes after the Beehive Indicator eruption, when the estimated constant added varied between +11.9 minutes to +14 minutes. The optimal constant was determined as +12 minutes following the eruption of Beehive Indicator, with a peak at 94.7% chance that the geyser will erupt during this time (see **Figure 3**). The percentage of the time each method (mean, median, mode, optimal constant) accurately predicted the geyser within a certain time range was calculated and shown in **Table 2** for each method ranging from a window of 2 minutes (+/- 1 minute) to a window of 30 minutes (+/- 15 minutes). The same target range (+/- 10 minutes) as Old Faithful (20-minute) was shown in bold. In the optimal constant estimate, 12 minutes were added to the eruption of Beehive Indicator to predict the eruption time of Beehive Geyser. Our results show that the National Park Service model of adding 17 minutes to the Beehive Indicator eruption time is significantly less accurate, being 9.4% less accurate than our optimal estimate.

**DISCUSSION**

The eruption times of two geysers, namely Old Faithful and Beehive Geyser, with available predictions from the Yellowstone National Park were analyzed in this paper. Our initial hypothesis prior to this study was that the predictions from the National Park Service for both geysers are likely to be accurate. However, our results demonstrated that the official predictions from the US National Park Service are variable in accuracy. In particular, the highly accurate predictions for Old Faithful from the National Park Service and our own regression models could be because it is one of the only geysers in the park that is constantly monitored, and the prediction algorithm is being adjusted constantly to ensure maximum precision. In contrast, the official predictions from the National Park Service for Beehive Geyser are significantly worse than our predictions that estimated the constant time to be added to the Beehive Indicator eruption time using the mean, median, mode or optimal methods.

We observed that the eruption time of Beehive Geyser is highly predictable when predicted using the duration of the Beehive Indicator eruption. Therefore, the best way to ensure being able to witness an eruption is to time the indicator eruption time, and if the eruption is more than 30 seconds, expect an eruption in around 12 minutes. On the other hand, since Beehive Indicator and Beehive Geyser are on the same boardwalk, our recommendation for tourists is go there immediately after the indicator goes off and wait for an eruption. In contrast, the National Park Service predictions generally overestimate the time interval between the Beehive Indicator and Beehive Geyser eruptions, which could lead to tourists arriving too late to see the beginning of an eruption.

The difference between the National Park Service model of adding 17 minutes to the Beehive Indicator eruption time and our optimal estimated constant of 12 minutes may be due to the definition of Beehive Geyser eruptions or inconsistent calculations of time intervals (whether counting started when Beehive Indicator erupted, or when it ended). This discrepancy could also be caused by predictions being linked to other factors such as the Beehive South Bubbler, however if there is a correlation between the eruption time and the Beehive South Bubbler, it remains undocumented on the National Park Service's website and is labeled as merely an object that bubbles during an eruption of Beehive Geyser.

This study focused on the Old Faithful and Beehive Geyser in the Upper Geyser Basin, as it is one of the more popular areas of Yellowstone. A limitation in the selection of geysers is the limitation of available data. GeyserTimes (11) is the only source, and it provides few attributes, generally only including a time stamp, eruption duration, and eruption height. While it appears that the Geyser Conservancy had a database that used to provide further data including the temperature and pH, it is no longer available for download and cannot be accessed online, thus vastly limiting the possible selection of geysers.

In terms of future work, analyzing additional geysers in the Upper Geyser Basin, such as the Castle and Grand Geyser, would further evaluate the accuracy of different predictive methods. With more data from other forms of indicators such as bubbling pools for Castle Geyser, refined predictions could facilitate people to enjoy the beauty of Yellowstone National Park. In addition, reducing the uncertainty of predicted eruption times would also improve the experience significantly. Currently most predictions are measured with an uncertainty of 20 minutes, however decreasing this could allow visitors to experience even more of the park, as well as reduce over-crowding and waiting times prior to eruptions.

Despite the extensive work undertaken to predict the eruption time of Old Faithful, there has been limited analysis of the water temperature and the height of the eruption. While the height of each eruption is sometimes recorded by the National Park Service, the current data collected is limited in availability and exhibits high uncertainty of 10 meters. The water temperature the water in the Old Faithful basin used to be recorded as well, however this data is no longer available from the National Park Service nor GeyserTimes. Both the height and temperature could be added as independent variables in regression models to further refine predicted eruption times of Old Faithful. Like Old Faithful, Beehive Geyser's analysis could benefit from incorporating additional variables in the regression models. Other geysers associated with the Beehive Geyser such as data from the Beehive Geyser South Bubbler could also be included in the analysis to generate even more accurate predictions for Beehive Geyser.

Another area of further work is to map out the tunnels underneath the geysers. Many of the geysers are connected underground (3) and eruption times of these connected geysers may exhibit certain patterns. Recently, Fagan et al. applied machine learning methods to study the pairwise interactions between geysers in the Upper Geyser Basin (14). Since challenges emerge as the paths are constantly changing (3), it may be also possible to use eruption data over time to learn how the underground interconnections may have changed, and thus, further advance the understanding of geysers as well as refinement of predictions.

Independent predictions from GeyserTimes (11) and many other apps on both the Google Play Store and the Apple App Store provide independent predictions that may be more accurate than the official predictions from the National Park Service. It is likely that the discrepancy in their predictions has been noticed. We hope that the US National Park Service will further refine their predictions, publish more public data to allow better analysis, and incorporate other independently published models into their predictions through wisdom of the crowds.

**MATERIALS AND METHODS**

The major steps of this study are summarized in **Figure 4**. The first two steps focused on data download, processing and filtering. The third and fourth steps focused on data analysis, visualization and statistical models. In the last step, the accuracy of my own predictions was compared to the published predicted and actual eruption times.

In the first step, all data for the geysers were retrieved from the GeyserTimes database (11). While having only one source of data increases uncertainty, it is a publicly available data source recommended by the National Park Service and consists of recent data for multiple geysers. In the second step, a Python script was written to systematically filter the data. The script removed incomplete and records that included missing data, such as the duration, as well as identified gaps in the data by using the National Park Service's range of times between eruptions to prevent the possibility of large gaps skewing the predictions.

**Data download and filtering for Old Faithful (steps 1 and 2)**

Using GeyserTimes (11), over 175,000 datapoints of Old Faithful eruptions from 1970 onwards were downloaded. Since the frequency of Old Faithful's eruptions has decreased over time due to a combination of earthquakes and vandalism (15), only data from 2010 onwards were used in this study. A typical interval between eruptions is expected to be in the range of 34 to 110 minutes, outlying datapoints and preliminary spurts (9) were removed by eliminating any eruptions that fell outside of this range. GeyserTimes included datapoints with variable precision in hours, minutes, or seconds. Therefore, datapoints measured in hours were removed, while datapoints in minutes and seconds datapoints were kept reducing uncertainty of measurements. These data filtering criteria in step 2 were implemented in a Python script. The resultant data consisted of 2029 separate eruptions from January 2010 to July 2022.

**Analysis for Old Faithful (steps 3 and 4)**

Software used included Word, Excel to create charts and filter data, and the Python programming language to reduce the size of the data and format the data into a spreadsheet format. The Python scripts used are available in our project GitHub repository at https://github.com/DanielRhee/Yellowstone-Geyser-Prediction.

The Old Faithful has been predicted as a bimodal geyser since 1959, meaning it generally erupts with two distinct durations (9). Old Faithful has been noted to exhibit correlation between the duration of an eruption and the time between eruptions (9). Three types of regression methods, namely, linear, exponential, and logistic, were used to generate predictions. Logistic regression was used because of the bimodal nature of Old Faithful making the eruption intervals categorical. In our regression models, the independent variable (x) is defined as the duration (difference between the end time and start time) of the current eruption). The dependent variable (y) is defined as the difference between the start time of the next eruption and the end time of the current eruption, as illustrated in **Figure 5**. The detailed predictions for each eruption versus actual eruption time are available as a spreadsheet on the GitHub at https://github.com/DanielRhee/Yellowstone-Geyser-Prediction. These regressions were calculated using an online Desmos graphical calculator.

**Evaluation of Old Faithful predictions (step 5)**

To evaluate the accuracy of these models, the percentage of the time a prediction was within $z$ minutes of the actual eruption time was calculated, where $z=1, 2, 3…, 15$. This method was chosen over more traditional methods such as the root mean squared error because predicting the eruption time precisely is unrealistic, and instead a range of time better represents the uncertainty in geyser eruption intervals.

**Data download and filtering for Beehive Geyser (steps 1 and 2)**

Similar to Old Faithful, data for Beehive Indicator and Beehive Geyser since 1970 were downloaded from GeyserTimes (11). Beehive is known to go dormant for some periods, where it did not erupt, meaning that there is missing data. Nearly 10,000 datapoints were downloaded. Unlike Old Faithful, Beehive Geyser is not a bimodal geyser, meaning that there is no consistent time frame for the time between eruptions, nor the time between the Beehive Indicator and Beehive Geyser eruptions. Thus, no datapoints were removed as outliers.

**Analysis for Beehive (steps 3 and 4)**

After downloading the eruption times of Beehive Geyser and Beehive Indicator, the difference between the Beehive eruption time and the indicator eruption times in minutes were calculated using Excel. These time differences were ordered against eruption numbers as shown in **Figure 2**. The mean, median and mode of the differences between Beehive eruption times and indicator times were computed across all eruptions to determine which method would be the best predictor of eruption time. The mean was 13.3 minutes, the median was 13 minutes, and the mode was 14 minutes. The mean, median and mode predicted eruption time of Beehive Geyser were calculated by adding 13.3 minutes, 13 minutes, and 14 minutes respectively to the Beehive Indicator eruption time. A spreadsheet showing these detailed results are available from our Github repository at https://github.com/DanielRhee/Yellowstone-Geyser-Prediction.

**Evaluation of Beehive predictions (step 5)**

Next, the accuracy of all differences between the Beehive eruption time and indicator eruption time were calculated in increments of 0.1 minutes using Excel. The percentage of the time each model (mean, median or mode) accurately predicted the geyser within a certain time range, from a window of 2 minutes (+/- 1 minute) to a window of 30 minutes (+/- 15 minutes), was calculated, as shown in **Table 2**.

**ACKNOWLEDGMENTS**

The authors would like to thank Ms. Maria Rallos at Interlake High School for her feedback.

**REFERENCES**

1. "Old faithful virtual visitor center: Upper geyser basin." https://www.nps.gov/features/yell/ofvec/exhibits/treasures/ugb/index.htm.
2. Bryan, T.S. *The geysers of yellowstone*. University press of Colorado, 2008.
3. " National park service: Hydrothermal features." https://www.nps.gov/yell/learn/nature/hydrothermal-features.htm.
4. "National park service: Steamboat geyser." https://www.nps.gov/yell/learn/nature/steamboat-geyser.htm.
5. "Montana state university: Beehive geyser." http://rcn.montana.edu/Features/Detail.aspx?id=9083.
6. Kuta, S. "Scientists map yellowstone's underground 'plumbing'." *Smithsonian Magazine*, edited by 2022.
7. Hurwitz, S. *et al.* "Why study geysers." *Eos*, vol. 102, no. 11, 2021.
8. "Old faithful: Beehive geyser." https://www.nps.gov/features/yell/tours/oldfaithful/beehive_work.htm.
9. Bauer, C.M. *et al.* "Old faithful, an example of geyser development in yellowstone park." *Yearbook of the Association of Pacific Coast Geographers*, vol. 5, no. 1, 1939, pp. 45-48.
10. "Old faithful virtual visitor center: Predicting geysers." https://www.nps.gov/features/yell/ofvec/exhibits/eruption/prediction/predict7.htm.
11. "Geysertimes." https://geysertimes.org/.
12. "Predicting geysers. Retrieved from geysertimes." https://geysertimes.org/predict/?aurum.
13. "The geyser observation and study association: Beehive geyser." https://web.archive.org/web/20201022100940/http://www.geyserstudy.org/geyser.aspx?pGeyserNo=BEEHIVE.
14. Fagan, W.F. *et al.* "Quantifying interdependencies in geyser eruptions at the upper geyser basin, yellowstone national park." *Journal of Geophysical Research: Solid Earth*, vol. 127, no. 8, 2022, p. e2021JB023749.

15. "Montana state university: Old faithful geyser." http://rcn.montana.edu/Features/Detail.aspx?id=9909.

**Figures and Figure Captions**

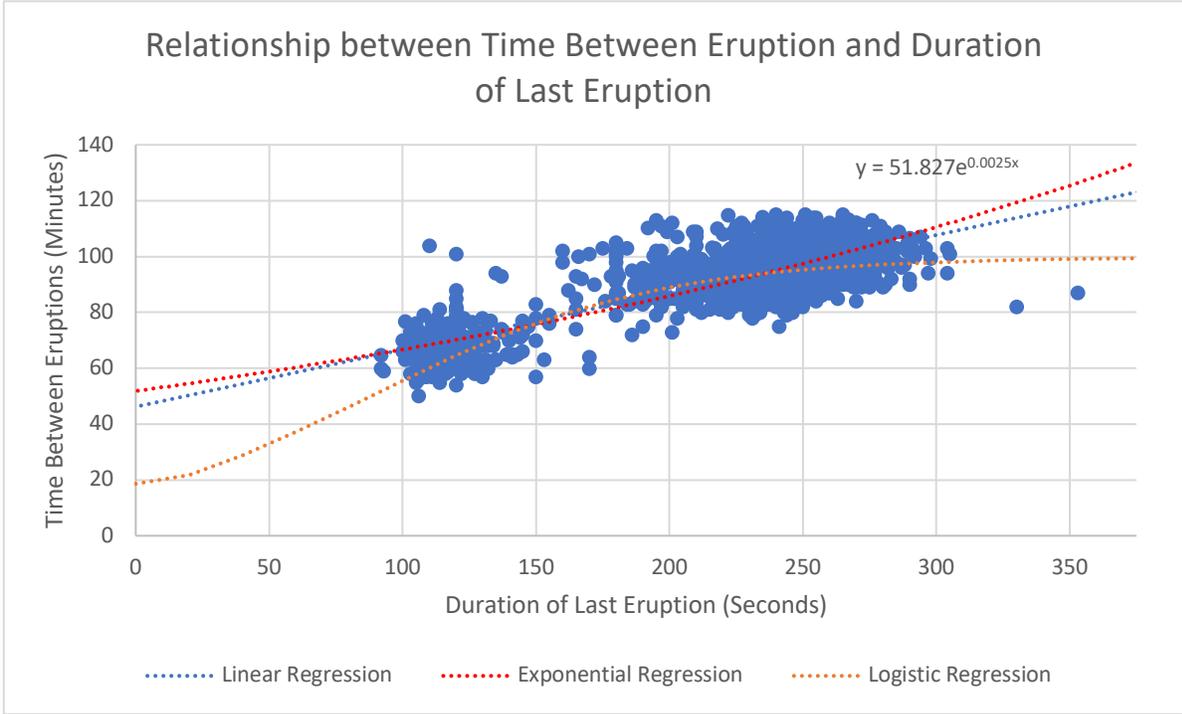

**Figure 1.** Relationship between time between eruptions and duration of the last eruption for Old Faithful.

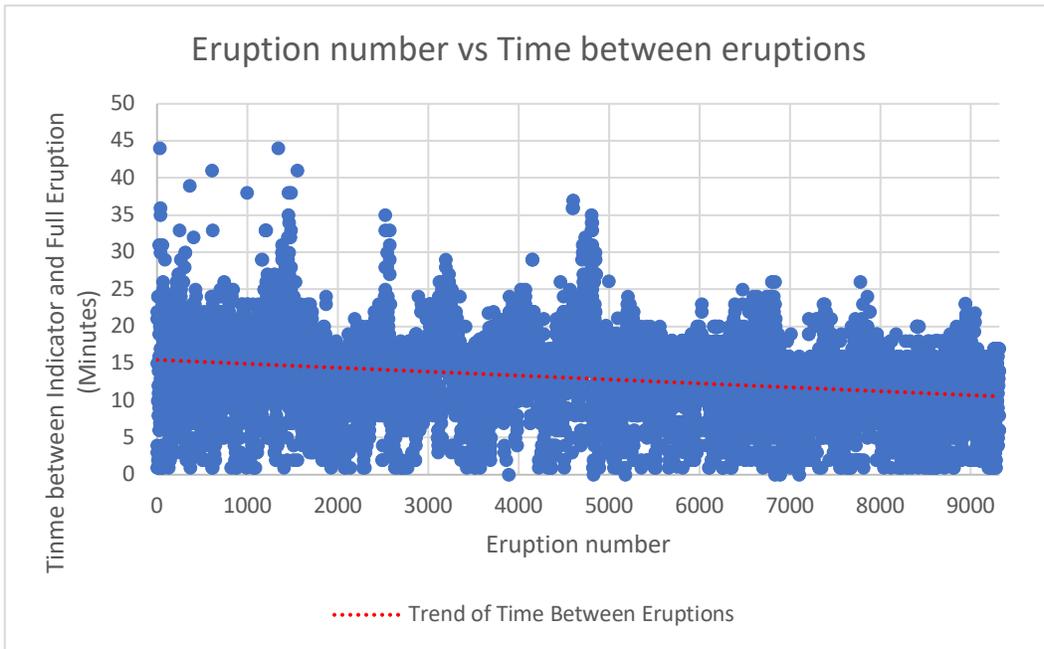

**Figure 2.** Plotting the difference between eruption time of Beehive Geyser and the eruption time of Beehive Indicator against the order of eruption.

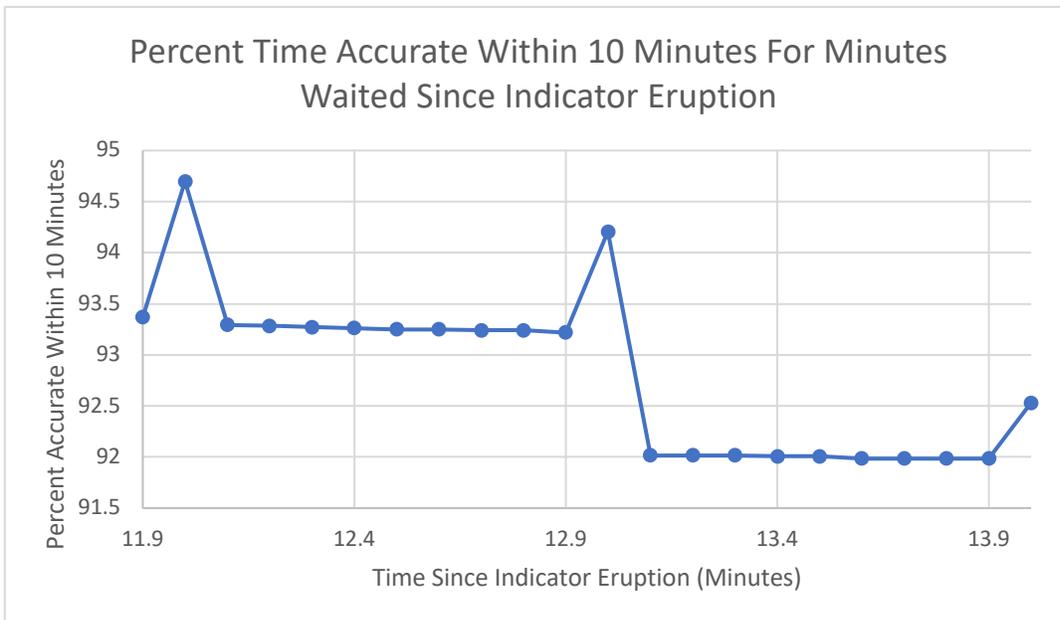

**Figure 3.** Percent accuracy for actual Beehive Geyser eruption times within 10 minutes after the Beehive Indicator erupted when the constant added varied between 11.9 to 14 minutes.

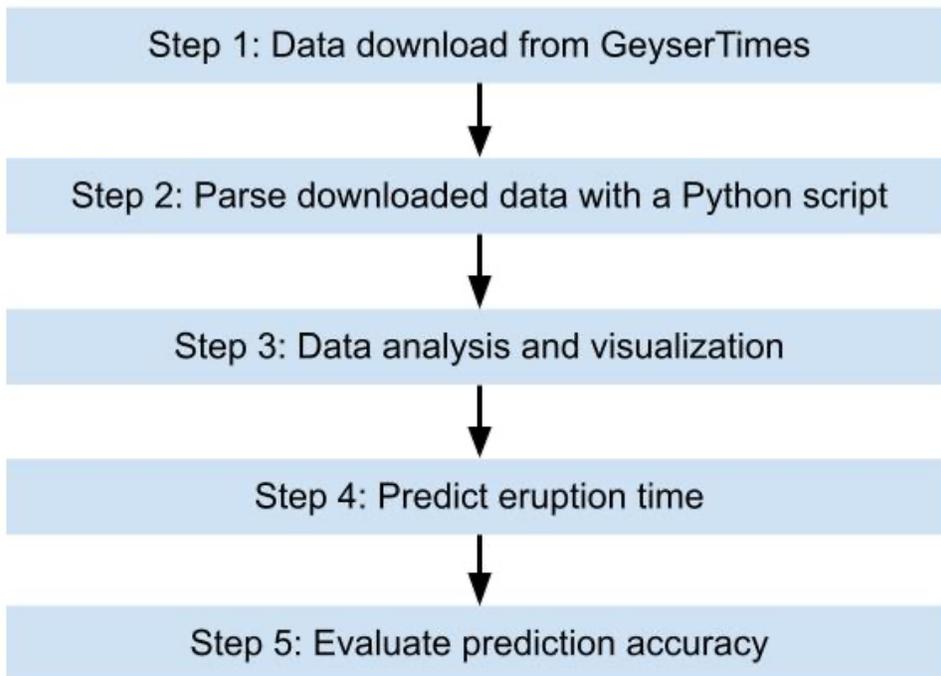

**Figure 4.** Summary of the steps of research procedure.

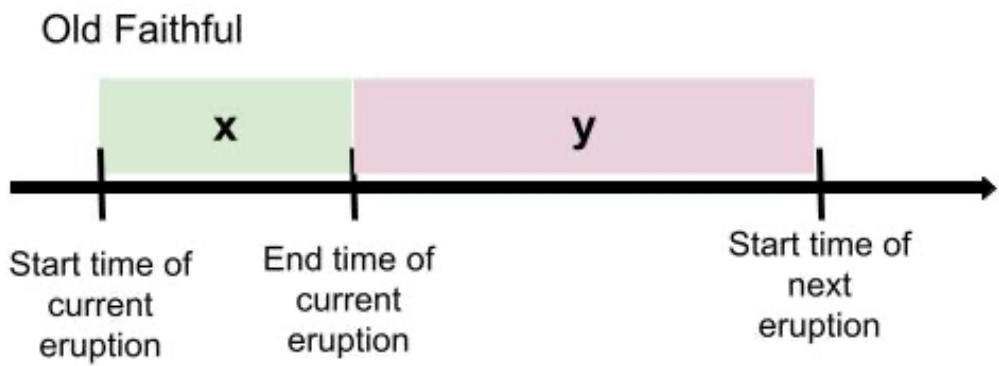

**Figure 5.** Illustration of regression models for Old Faithful. The independent variable (x) is defined as the duration of the current eruption. The dependent variable (y) is defined as the difference between the start time of the next eruption and the end time of the current eruption.

**Tables with Captions**

| +/- Amount of minutes in each direction | Percent accurate Within given min (Linear) | Percent accurate Within given min (Sigmoid) | Percent accurate Within given min (Exponential) | Percent accurate Within given min (NPS) |
|---|---|---|---|---|
| 1 | 12.4 | 12.9 | 5.57 | 15.5 |
| 2 | 22.8 | 24.9 | 22.6 | 26.2 |
| 3 | 33.6 | 35.2 | 33.7 | 36.7 |
| 4 | 44.2 | 45.6 | 42.9 | 46.7 |
| 5 | 52.8 | 53.8 | 52.0 | 56.1 |
| 6 | 60.1 | 62.5 | 59.5 | 63.5 |
| 7 | 68.3 | 70.3 | 67.4 | 70.9 |
| 8 | 74.5 | 76.7 | 73.4 | 76.9 |
| 9 | 80.2 | 82.6 | 79.0 | 82.0 |
| **10** | **84.9** | **86.1** | **84.0** | **86.1** |
| 11 | 89.3 | 89.2 | 87.6 | 88.9 |
| 12 | 91.9 | 92.1 | 90.9 | 91.1 |
| 13 | 94.2 | 94.0 | 93.2 | 92.9 |
| 14 | 95.3 | 95.4 | 94.9 | 94.5 |
| 15 | 96.5 | 96.4 | 96.0 | 95.9 |

**Table 1.** Comparing regression model accuracy within a certain time interval, ranging from 2 minutes (+/- 1 minute) to 30 minutes (+/- 15 minutes). The US National Park Service (NPS) target uncertainty range (+/- 10 minutes) is shown in bold.

| +/- Amount of minutes in each direction | Percent accurate Within x min (Mean) | Percent accurate Within x min (Median) | Percent accurate Within x min (Mode) | Percent accurate Within x min (Optimal) | Percent accurate Within x min (NPS) |
|---|---|---|---|---|---|
| 1 | 0.158 | 0.215 | 0.221 | 0.197 | 0.174 |
| 2 | 0.306 | 0.356 | 0.365 | 0.333 | 0.287 |
| 3 | 0.441 | 0.483 | 0.487 | 0.450 | 0.402 |
| 4 | 0.552 | 0.581 | 0.597 | 0.560 | 0.500 |
| 5 | 0.646 | 0.674 | 0.673 | 0.656 | 0.589 |
| 6 | 0.718 | 0.748 | 0.741 | 0.746 | 0.666 |
| 7 | 0.785 | 0.813 | 0.804 | 0.807 | 0.728 |
| 8 | 0.843 | 0.863 | 0.856 | 0.861 | 0.773 |
| 9 | 0.885 | 0.904 | 0.894 | 0.911 | 0.813 |
| **10** | **0.920** | **0.942** | **0.925** | **0.947** | **0.853** |
| 11 | 0.951 | 0.968 | 0.955 | 0.973 | 0.889 |
| 12 | 0.974 | 0.985 | 0.977 | 0.981 | 0.916 |
| 13 | 0.989 | 0.989 | 0.991 | 0.985 | 0.940 |
| 14 | 0.992 | 0.991 | 0.992 | 0.989 | 0.965 |
| 15 | 0.992 | 0.992 | 0.994 | 0.991 | 0.984 |

**Table 2.** Percentage accuracy within a time range for each prediction method for the eruption time of Beehive Geyser by adding the computed mean, median, and mode to the eruption time of Beehive Indicator. The US National Park Service (NPS) target range is (+/- 10 minutes) is shown in bold.